# A Systematic Literature Review on Persuasive Technology at the Workplace

Kilian Wenker
Friedrich-Alexander-Universität (FAU) Erlangen-Nürnberg

**Abstract.**
Employees face decisions every day — in the absence of supervision. The outcome of these decisions can be influenced by digital workplace design through the power of persuasive technology. This paper provides a structured literature review based on recent research on persuasive technology in the workplace. It examines the design and use of persuasive systems from a variety of disciplinary perspectives and theories. The reviewed studies were categorized into the research streams of technology design, user-centered research, and gamification. The purpose of the studies is categorized using a modified definition of the persuasive systems design model. A number of experimental studies show that alignment of the employee's behavior with the employer's agenda can be achieved. A robust finding is the key role of interactivity in granting employees a subjective experience of rapid and meaningful feedback when using the interface.

# 1 Introduction

Persuasive Technology (PT) refers to interactive information technology designed for changing users' attitudes or behavior in the field of human-computer interaction (HCI) [1]. With its manipulative and often invisible power to influence employees, it might be game-changing for digital workplace design.

Take Microsoft as an example, where Ross Smith leads a team of testers who put unified communications products through the paces to find defects. When it was hard to find enough employees around the world willing to review Windows dialog boxes in their spare time, Ross Smith invented a language quality game and made employees compete against each other to win the most points — making employees go above and beyond their work responsibilities [2, 3].

Incentives work both ways, with punishment instead of a positive reward. Take, for example, a medium-sized call center in Germany with several branches that take inbound telephone calls for customer service and sales. If a customer wants to cancel his subscription, the employees shall dissuade the customer from unsubscribing in a so-called retention attempt. A discount may be offered for this purpose as a last resort. However, many employees tend to give this discount liberally since it increases their customer retention rate and simplifies the phone call. This in turn hurts the company's profit margin. The work flow software has been rewritten to introduce a random chance of blocking the discount. This makes the employee wait until the very end before offering a discount. What happens here is ultimately an automated, somewhat forced change in behavior, because an employee promising the discount without actually being able to grant it at the end of the call, has a much more difficult conversation afterwards with the customer who will certainly quit after a disappointed expectation. Here, PT comes with discouragement and repetition. Discouragement in the form of harmful consequences as opposed to reward. Repetition as an effective instrument of persuasion.

PT creates opportunities for various organizational needs. It may persuade customers if you consider one-click checkouts or purchase recommendations on websites like amazon.com. Or it may help managers to persuade employees like in the two workplace examples above. My research goal is to *provide an overview and classification of recent research on PT at the workplace*. To reach this goal, I perform a literature review following the principles laid down by Webster and Watson [4]. The rest of this paper is organized as follows; Section 2 establishes the theoretical background underlying my systematic review; Section 3 explains the literature review process and Section 4 summarizes the main findings thereof; Section 5 discusses the implication of those findings. Finally, the paper is concluded in Section 6.

# 2 Theoretical Background

Workplace design has significantly changed since human work has been increasingly disrupted or even determined by information and communication technology. A HCI environment necessitates a deep understanding of various disciplines, since designing digital technology for a group of employees involves psychology (how humans behave as individuals), sociology (how humans behave in a group), computer science (how to develop software), organizational studies (how organizational structures, processes, and practices work), and information systems (how to develop, use, and apply information technology in the business) [5].

The term persuasion refers to an attempt to shape, reinforce, or change behaviors, feelings, or thoughts about an issue, object or action [1]. PT is broadly defined as technology that aims to change user behaviors or underlying attitudes [1, 6–8]. This concept has recently been challenged, because PT raises questions around the borderlines between encouragement, persuasion, and in particular coercion [9]. Advocates of the classical definition (which in general excludes coercion) suggested the term behavior change support systems [10]. Other researchers started explicitly avoiding the term PT and suggested using behavior change technologies instead, explicitly including coercion [11]. This paper will include coercion to incorporate a setting such as the call center mentioned in the introduction, but will stick with the term PT.

As might be expected from an interdisciplinary field like HCI, the research has split in two directions: the technological design which focusses on the technological aspects, and on the other hand user-centered research which encompasses the human aspects [12]. Most research focusses on the latter. Figure 1 depicts how the amount of writing on PT has been increasing steadily. The graph shows that about ten years ago, the number of studies on PT first began to stagnate. What stands out in this chart is a sharp rise in the number of studies related to gamification, at around the same time.

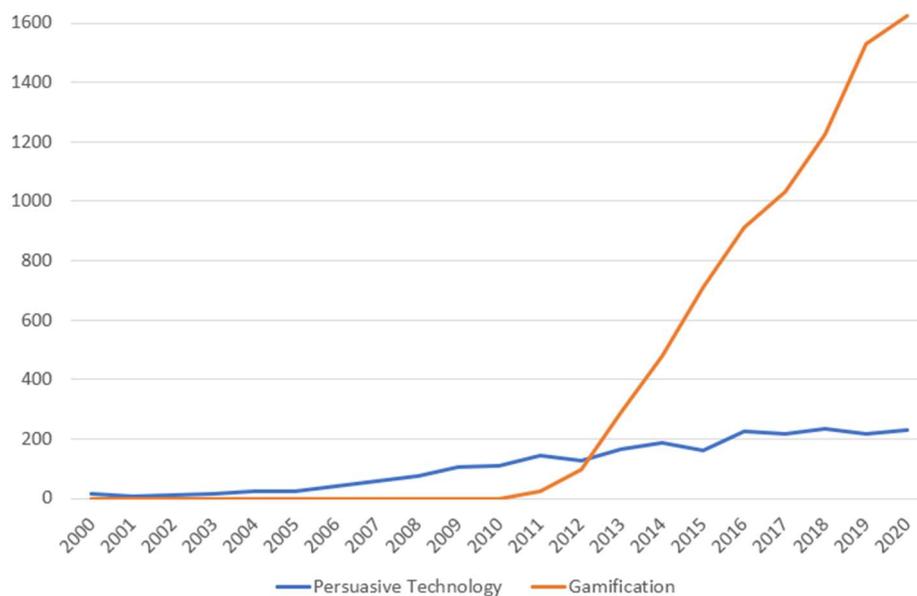

**Fig. 1.** Number of search hits, by year, for the queries (TITLE-ABS-KEY(gamification) AND PUBYEAR > 1999) and (TITLE-ABS-KEY(Persuasive technology) AND PUBYEAR > 1999) in the Scopus database. Adapted from Hamari et al. [7]

One reason for this development was that PT failed regarding a practical implementation in companies. This can be seen in the cases of Disneyland and Paradise Pier Hotels, where public monitors displayed leaderboards showing efficiency numbers in green for the quickest employees and in red for others, in order for housekeepers to become more efficient [13]. Many employees felt they were being controlled and even referred to this as an electronic whip. At the same time, gamification became increasingly popular as a successful approach. Gamification is generally understood as the use of game design elements and principles to make everyday tasks more engaging — but in non-gaming contexts [13, 14]. Thus, it's part of the definition of PT that this paper uses, if technology is involved. A more nuanced view highlights that if employees cannot turn off the game, then it's no longer a game [15].

Previous research on PT has emphasized the importance of designing persuasion into technologies. The persuasive systems design (PSD) model represents the main conceptual framework on PT [8–10, 16]. It was specifically developed for HCI and distinguishes three categories, namely understanding key issues behind persuasive systems, analyzing the persuasion context and design of system qualities. These three categories collect various elements, which are interrelated, overlap, and are formulated in very general terms.

Leaving the technological aspect aside, behavioral theory inevitably comes to the fore. The theory of planned behavior (TPB) is a psychological theory that states that three interrelated core factors, namely attitude, subjective norms, and perceived behavioral control, determine human intentions. Learning theory seeks to describe how students receive, process, and retain knowledge, influenced by cognitive, emotional, and environmental influences, as well as prior experience. In consequence, learning is considered to be an aspect of conditioning and a system of rewards and goals would be helpful. Further prominent theories related to behavior change include the theory of reasoned action, the technology acceptance model, the self-efficacy theory, social cognitive theory (SCT), the elaboration likelihood model, the cognitive dissonance theory, the goal setting theory and computer self-efficacy [11, 17–19] (see [10] for an overview of these and further theories).

## 3   Research Method

The review process began with the selection of the data bases to be used for the literature searches. I chose the databases *Scopus*, *Business Source Complete* and *Web of Science*. For *Scopus*, I conducted a keyword search formulating the following query: (TITLE-ABS-KEY( persuasive AND systems AND behavior ) OR TITLE-ABS-KEY( behavior AND change AND support) OR TITLE-ABS-KEY( digital AND work AND design AND behavior)) AND PUBYEAR > 1999 AND ( LIMIT-TO ( SUBJAREA, "COMP") OR LIMIT-TO ( SUBJAREA, "ENGI") OR (SUBJAREA, "SOCI") OR LIMIT-TO ( SUBJAREA, "BUSI" ) OR LIMIT-TO ( SUBJAREA, "PSYC") OR LIMIT-TO ( SUBJAREA, "MULT")) AND ( LIMIT-TO (LANGUAGE, "English") OR LIMIT-TO ( LANGUAGE , "German")). I performed the search in May 2021, which resulted in more than 82,000 hits. Next, I customized those results with a post-query filter by filtering the articles which have been cited at least 150 times. That second step reduced the remaining hits drastically to 171.

The second database I queried is *Business Source Complete*, using the following search terms "work design" AND "persuasive". Those were limited to "Peer Reviewed Journals" and narrowed by subject "persuasive technology". The search mode I applied was "SmartText Searching". Additionally, I expanded the query using internal synonyms to include "equivalent subjects". From this, *Business Source Complete* returned 19 results.

I used a third database, *Web of Science*, which focusses on technical aspects. Accordingly, it provided a low coverage of four articles in total, all of which could be ruled out after close inspection.

Next, I removed duplicates (from the first two search systems) and filtered the titles and abstracts for relevant papers, which resulted in 33 hits. Full-text analysis served to assess the content of the remaining articles.

Quality assessment of the 33 primary studies eligible for review was undertaken in parallel with the creation of a concept matrix.

Figure 2 illustrates how 196 papers fit the initial inquiry, and how that number decreased to the final sample of 19 papers.

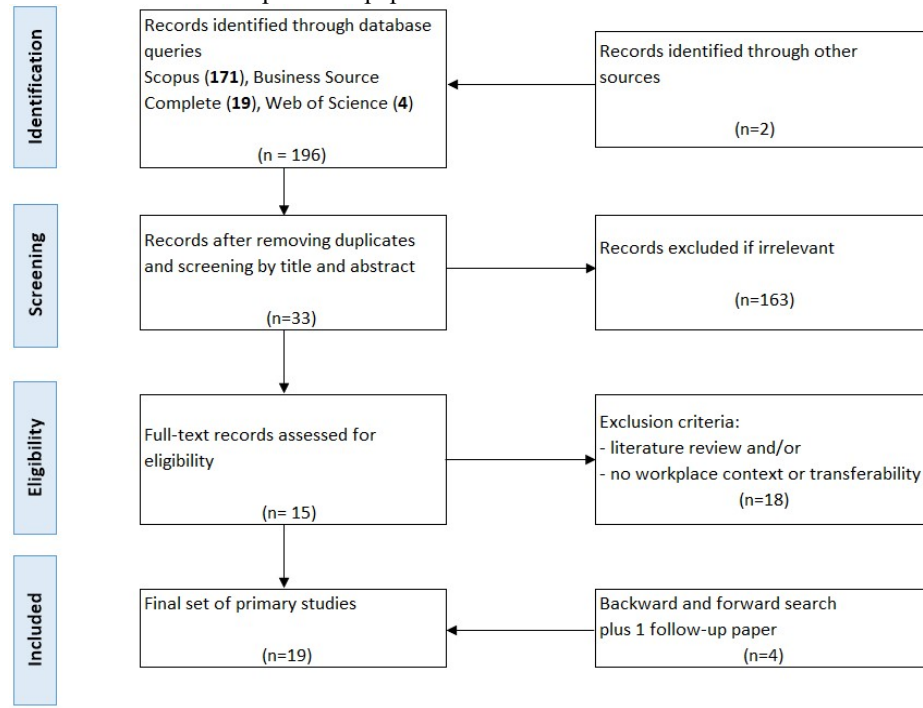

**Fig. 2**. Flow diagram for study selection. Adapted from Johnson et al. [20]

Using backward citation searching, I found a paper [21] describing the development of a social media application that promotes energy awareness while motivating employees to engage in energy-saving behaviors. In the follow-up publication [22], usability tests were conducted with 128 subjects as proxies for typical office building users. Since the results of that study were published across two separate research papers, I considered them to represent one primary study.

To classify the *content of research*, I will allocate the three historic *research streams* of technological design, user-centered research, and gamification to the general direction of the research reviewed. Another way to break down the content is to draw a distinction between the three phases of the PSD framework. But since the three terms are more like catalog headings, I decided to redefine them to reflect the *purpose of the study* as follows. *Understanding key issues*: These papers aim to understand the fundamentals behind persuasive systems, whether they are technical or user-centered. *Analyzing the persuasion context*: The topic of papers in this category gravitates on a concrete, working or applied PSD. The focus may be analyzing the PSD in its entirety or investigating parts of it like the persuader, the persuadees, the intent, the strategies, technology dependent features or other aspects. *Design of system qualities*: While research of the second category provides information on content and functionality, there is a growing body of literature that recognizes the further development of existing possibilities. This research focusses on improving or evaluating an existing PSD.

A second concept matrix will assess the *scientific approach* of the reviewed papers, distinguishing studies focusing conceptual work, quantitative research and qualitative research.

I classify the *research design* into lab experiments, crowdsourcing, in-field experiments and automated evaluation procedures (based on [23]).

As the name implies, *post-hoc explanations* represent an attempt to make sense of the collected data after the fact. Statistical analysis has been instrumental in our understanding of the interdependence of actions, cognitions and emotions. However, it is hard to tell what exactly is cause and effect. Observational or descriptive explanations strictly adhere to the data and refrain from interpreting what might be cause or effect. Although presume and assume both mean to assume something to be true, an assumption suggests there is little or no evidence supporting your guess, and a presumption implies more confidence or evidence backed reasoning [24]. In this paper, I use the term explanatory assumptions if authors suppose a relation or conclusion without providing evidence.

## 4 Results

The studies represent a range of different domains, including theoretical frameworks, empirical studies and four case studies. The most widely represented domain was sustainability at the workplace (including energy saving). One study is notable for the unique persuasive approach of negative emotion, i.e. perceived threat from malicious IT [19]. Most studies focus on the individual and describe the work setting, but include little context on the technology itself. Table 1 displays a concept matrix with all identified concepts for the content of the researched studies — the headings for those concepts being research stream and purpose of the study as presented in Section 2.

**Table 1.** Concept matrix for the content of the reviewed literature

| Articles and reference | | Research stream | | | Purpose of study | | |
|---|---|---|---|---|---|---|---|
| | | Technology design | User-centered research | Gamification | Understanding key issues | Analyzing the persuasion context | Design of system qualities |
| Johnston and Warkentin. (2010) | [19] | • | • | | • | | |
| Hekler et al. (2013) | [11] | | | | • | | |
| Kumar (2013) | [25] | • | • | • | • | • | |
| Makanawala et al. (2013) | [26] | | | • | | • | |
| Oinas-Kukkonen (2013) | [10] | • | • | | • | • | |
| Whitson (2013) | [15] | • | • | • | • | • | |
| Fritz et al. (2014) | [27] | | • | | | • | |
| Lehrer et al. (2014) | [22] | • | • | • | • | • | |
| Lockton et al. (2014) | [28] | | • | • | | • | |
| Hamari and Koivisto (2015) | [18] | | • | • | • | | |
| Filippou et al. (2016) | [6] | | • | | | • | |
| Chesney et al. (2017) | [29] | | • | | • | | |
| Liu et al. (2017) | [13] | | | • | • | • | |
| Stieglitz et al. (2017) | [30] | | • | | | | |
| Mogles et al. (2018) | [31] | | • | | • | • | |
| Khashe et al. (2019) | [32] | | • | | | • | |
| Böckle et al. (2020) | [12] | | • | • | • | | • |
| Chiu et al. (2020) | [17] | • | • | | | | • |
| Khataei et al. (2021) | [33] | • | • | | • | | |

The studies represent a range of scientific approaches, including theoretical work, quantitative research, qualitative research, and a mixture thereof. Some theoretical, conceptual papers did not include any empirical data. The most popular discipline was quantitative research nonetheless. A common feature across many articles was the limited amount and quality of empirical evidence, often acknowledged by the authors themselves in the disclosed limitations of their studies. Table 2 displays a concept matrix with all identified concepts. These concepts are summarized under the headings scientific approach, research design and post-hoc explanation (of results), as presented in Section 3.

**Table 2.** Concept matrix for the scientific methodology of the reviewed literature

| Articles and reference | | Scientific approach | | | Research design | | | | Post-hoc explanation | | | |
|---|---|:---:|:---:|:---:|:---:|:---:|:---:|:---:|:---:|:---:|:---:|:---:|
| | | Conceptual work | Quantitative research | Qualitative research | Lab experiments | Crowdsourcing | In-field experiments | Automated eval. procedures | Statistical analysis | Observational and/or descriptive | Explanatory presumptions | Explanatory assumptions |
| Johnston and Wa. (2010) | [19] | • | • | | | | • | | • | | • | |
| Hekler et al. (2013) | [11] | • | | | | | | | | | | |
| Kumar (2013) | [25] | • | | | | | | | | | | |
| Makanawala et al. (2013) | [26] | | | • | | | | | | | | |
| Oinas-Kukkonen (2013) | [10] | • | | | | | | | | | | |
| Whitson (2013) | [15] | | | • | | | • | | | | | |
| Fritz et al. (2014) | [27] | | | • | | | • | | | • | | |
| Lehrer et al. (2014) | [22] | | • | • | • | | | | • | • | • | |
| Lockton et al. (2014) | [28] | | • | • | | | • | | • | • | • | |
| Hamari and Ko. (2015) | [18] | • | • | | | | • | | • | | • | |
| Filippou et al. (2016) | [6] | | • | • | | | | | | | • | • |
| Chesney et al. (2017) | [29] | | • | | • | | | | • | • | | |
| Liu et al. (2017) | [13] | • | | | | | | | | | | |
| Stieglitz et al. (2017) | [30] | | | • | | | | • | | • | • | |
| Mogles et al. (2018) | [31] | • | • | | • | | • | • | • | | • | |
| Khashe et al. (2019) | [32] | | • | | • | • | | | • | | • | |
| Böckle et al. (2020) | [12] | | • | | | • | | | • | | • | |
| Chiu et al. (2020) | [17] | | • | | | | • | • | • | • | | |
| Khataei et al. (2021) | [33] | | • | | | | • | | • | | • | |

 Research on PT is scant in workplace contexts and offers an excellent opportunity for future research. Therefore, I looked for PT in other settings if those studies were fundamentally appropriate to understand persuasion in the workplace. For instance, in a study on nudging people into eco-friendly behavior, 30 students worked on computers without change in their usual computer usage, but their computers' electricity consumption was measured and eco-feedback employed to persuade [17]. There is no obvious reason why the basic causalities should work differently for employees at PCs than for students at PCs. Other researchers extend this view by using mainly undergraduates for a simulation and stating that those students are proxies for typical office building occupants [22]. Researchers have been reflective and receptive to discussions on the wide use of students as experimental subjects and online panel data, gauging experiments against real-world behavior [34, 35]. In figure 2, I refer to proxy samples as transferability (to the workplace area).

Four case studies deal with companies that use PT in different areas of their business. The UK government's Department of Energy and Climate Change (DECC) introduced PT and some gamification mechanics to make workplaces more sustainable and succeeded e.g. in reducing the size of office space by encouraging employees to move into a common office space at night [28]. Another case study illustrates approaches to improve productivity, spirit, and engagement at work of customer service agents by introducing gamification into SAP Service OnDemand [26]. An analysis of the results of PT and gamification in call centers explores why the latter work or do not work there [15]. Finally, semi-structured interviews are used to gauge the impact of digital nudges at an automotive supplier [30].

## 5   Discussion

**Principles and strategies hide the lack of comprehensive models or theories.** I refer to the latter as a comprehensive and well-substantiated explanation of phenomena like PT. From my sample of 19 articles, I identified frameworks that mainly expose strengths and weaknesses of existing theories and incorporate those into a more or less structured theoretical framework [13, 16, 31] or compile a concise and quite comprehensive table of theories, listing only theories developed in social psychology, except for computer self-efficacy [10]. Two information systems theories explicate the cause behind actions and behaviors [13]. Media characteristics is concerned with users' choice of a type of medium. In SCT, any observed behavior can change a person's mindset (cognition), and human behavior is explained as a continuous, two-way interaction between cognitive, behavioral, and environmental influences.

The PSD framework is receiving attention as a key paradigm for PT research [10]. But it does not explain the causal relationships, it essentially lists 28 principles or strategies, leaving it unclear how these should be implemented in a particular user context [31]. Nevertheless, I found the rough breakdown useful and, after modification, I was able to use it to narrow down content concepts.

Most studies do not try to elaborate a comprehensive PT model or theory, but rather list specific principles or strategies, often in a selective manner (see [1] for an inventory of those principles or strategies). In summary, no consensual comprehensive theory has emerged so far [18]. Authors typically assemble an eclectic mix of principles, taken from different theories, to design a persuasive system rather than design a comprehensive theory. And since, for instance, a system designed to assist in weight loss is not easily adaptable to convince people about energy conservation [33], one could argue a sound, generalized model would have to incorporate too many aspects. In other words, the workings of behavior change are highly complex. Yet it is clear that the success of simple principles and strategies is what makes PT so appealing.

**Interactivity increases the persuasive effects of PT.** The term interactivity is used to describe the communication process that takes place between humans and computer software, for instance if a computer-based system expects a response from the user, and provides feedback to that response. But to see user-system interactions as something that the user does, represents a rather limited view given the level of interactivity involved in current systems [13]. The challenge of technology here is to provide a concrete cue for the behavior of interest, although a key question remains as to what form that cue should take, e.g. a motivational message, encouragement, reward, behavior prompts or reminders or any other persuasive communication method [31].

The literature affirms interactivity's ability to create a highly engaging and cognitively engaging experience. According to the TPB, behavioral intentions are affected by attitude, subjective norms, and perceived behavioral control while the comprehensive action determination model (CADM) points out that habits, intentions, and situations directly affect individual behavior [17].

Persuasive strategies such as competition, self-monitoring and feedback, goal-setting and suggestion, personalization, reward, social comparison lead to different effects in distinctive situations and may drift depending on mood and situation of the user's personality [12, 33]. An interactive social dialogue encourages users to comply more with requests when they otherwise might be more reluctant, a strategy that works differently across different types of people [29, 32]. Suggesting that interactivity plays a key motivational role in helping employees adjust their behavior is not new. What is new is how interactivity in the context of PT accelerates and amplifies this feedback by enabling rapid data collection and analysis, and a greater understanding and appreciation of the persuasive principles that shape human behavior. PT not only delivers quick and persuasive messages to the employee. Moreover, it creates an environment for the adoption of a particular action, i.e., the very act of *using* interactive PT privileges some options over others. There is a notable difference between, for instance, seeing and reading a generic advertisement on a website, a customized online advertisement based on browsing history and social media excerpts [33], and an interactive PT like a design pattern on a cookie consent request that nudges or even forces the user through a persuasive interface design.

Providing real-time feedback about employees' actions by amassing large quantities of data and then simplifying this data into modes that are easily understandable, such as progress bars, graphs and charts may help employees or managers in gauging the situation. However, one potential source of bias is surveillance. Just because a PT is successful in gamified spaces does not mean that the same technology will be equally accepted and treated in other spaces. For gamification mechanics [25, 28] to be experienced as a fun way to work, all employees need to be willing participants [15].

**Positive incentives are much more common than negative incentives.** A key aspect for the application of PT in the workplace is to cause desired employee behavior. Several theoretical models specify the motivational and cognitive antecedents of behavior change. For example, the theory of reasoned action [8, 10, 17, 19], the PBT [7, 10, 11, 17–19], or the theory of self-determination (SDT) [11–13, 17–19, 23, 31];

see [36] for an overview of how job motivation is affected by technologies and [13] for different theoretical perspectives on gamification research in the workplace.

Relying on SDT, and the distinction between intrinsic motivation (the drive to do something without external rewards and for its own sake) and extrinsic motivation (performing an activity to attain some separable outcome), the PSD framework proposes that rewarding target behaviors reinforces those behaviors and can increase the persuasiveness of a system, but it is the combination of rewards, collaboration, and competitive setting that is crucial [8, 13, 20]. There is evidence of a strong relationship between positive external rewards and certain behavior outcomes [20, 22, 25, 26]. While PT without gamification elements aims to support decision-making through cognitive processes (i.e., the reward is extrinsic), gamification, on the other hand, draws on affective processes, i.e., gamification is aimed at invoking employees' or users' intrinsic motivations through design reminiscent from games [18, 25]. If we look at gamification from the perspective of the traditional reward dichotomy, it would be difficult to categorize it as either intrinsic or extrinsic, since gamification provides both benefits: an external benefit such as task completion and an internal, hedonistic benefit such as fun. It is notable that making the consequences of a certain action visible to the employee, even in real-time PT, is not intrinsically meaningful information in itself. Employees would need a context like comparing the results to an objective or other employees' results [30], or personalized incentive mechanisms [12].

Negative incentives do exist but are often hidden from sight. Setting challenges and goals can be associated with different outcomes — positive if the employee succeeds or negative if the employee fails to reach the goal and that failure is measured and displayed.

While many studies focus on positive rewards, the most notable exception explores user behavior in reply to negative rewards: Malicious threats concerning IT security in the workplace inspire different outcomes for different users based on their perceptions of efficacy and threat, but combining threat severity with users self-efficacy and perceptions of response efficacy leads to more persuasive impact [19].

Irrespective of the above, positive rewards may not have the desired effect on employee behavior. A misaligned reward system can lead to an employee gaming the system, at the expense of the employer [6, 13, 27]. The same is true if system-based rewards exert a strong influence on people's personal goals, to such an extent that the system goals seemed to supplant underlying goals, i.e. the artificial rewards gets more important than the real task [6, 27].

## 6 Conclusion

The literature on PT strategies and principles continues to grow rapidly. Drawing upon that literature, this survey presented key concepts related to PT in the workplace. Although there does not appear to be serious disagreement among researchers about the motivational and cognitive antecedents of behavior change, the findings highlight a gap in comprehensive theory building. Interactivity facilitates feedback for employees by providing quick and compelling messages and also by requiring the employee to respond to the PT. Since employees are receptive to rewards and competition in the workplace, the results of this review imply that positive incentives are mostly successful in motivating and engaging employees, especially if feedback is meaningful and personalized. The stimulus is transmitted in part through cognitive processes and, in the case of gamification mechanisms, through affective processes.

Since the reviewed studies vary in their methods and in the details of the research questions, and lack a comprehensible view, researchers should address this issue by integrating from a diverse set of technical, behavioral, organizational, and social disciplines (to name a few). Moreover, employee preferences should be addressed, since empirical studies in personalization imply that the effects of incentives may be considerably different than predicted by generic theories [12, 33, 36].

# References

1. Fogg, B.: Persuasive Computers: Perspectives and Research Directions. In: Karat, C.-M., Lund, A., Coutaz, J., and Karat, J. (eds.) Proceedings of the SIGCHI Conference on Human Factors in Computing Systems - CHI '98, Los Angeles. pp. 225–232. ACM Press/Addison-Wesley Publishing Co., New York (1998). https://doi.org/10/ddts6d.
2. Werbach, K., Hunter, D.: For the Win: How Game Thinking Can Revolutionize Your Business. Wharton Digital Press, Philadelphia (2012).
3. Microsoft's Ross Smith asks shall we play a game?, https://blogs.microsoft.com/ai/microsofts-ross-smith-asks-shall-we-play-a-game/, last accessed 2021/07/05.
4. Webster, J., Watson, R.: Analyzing the Past to Prepare for the Future: Writing a Literature Review. MIS Quarterly. 26, xiii–xxiii (2002).
5. Richter, A., Heinrich, P., Stocker, A., Schwabe, G.: Digital Work Design: The Interplay of Human and Computer in Future Work Practices as an Interdisciplinary (Grand) Challenge. Business & Information Systems Engineering. 60, 259–264 (2018). https://doi.org/10/ghksjr.
6. Filippou, J., Cheong, C., Cheong, F.: Modelling the Impact of Study Behaviours on Academic Performance to Inform the Design of a Persuasive System. Information & Management. 53, 892–903 (2016). https://doi.org/10/gj44qz.
7. Hamari, J., Koivisto, J., Pakkanen, T.: Do Persuasive Technologies Persuade? - A Review of Empirical Studies. In: Spagnolli, A., Chittaro, L., and Gamberini, L. (eds.) Persuasive Technology. PERSUASIVE 2014, Padua. pp. 118–136. Springer International Publishing, Cham (2014). https://doi.org/10/ggsmqn.
8. Oinas-Kukkonen, H., Harjumaa, M.: Persuasive Systems Design: Key Issues, Process Model, and System Features. Communications of the Association for Information Systems. 24, (2009). https://doi.org/10/dk7n.
9. Purpura, S., Schwanda, V., Williams, K., Stubler, W., Sengers, P.: Fit4Life: The Design of a Persuasive Technology Promoting Healthy Behavior and Ideal Weight. In: Karat, C.-M., Lund, A., Coutaz, J., and Karat, J. (eds.) Proceedings of the SIGCHI Conference on Human Factors in Computing Systems - CHI '11, Vancouver. pp. 423–432. Association for Computing Machinery, New York (2011). https://doi.org/10/dt74wm.

10. Oinas-Kukkonen, H.: A Foundation for the Study of Behavior Change Support Systems. Personal and Ubiquitous Computing. 17, 1223–1235 (2013). https://doi.org/10/f482m8.
11. Hekler, E., Klasnja, P., Froehlich, J., Buman, M.: Mind the Theoretical Gap: Interpreting, Using, and Developing Behavioral Theory in HCI Research. In: Bødker, S., Brewster, S., Baudisch, P., Beaudouin-Lafon, M., and Mackay, W. (eds.) Proceedings of the SIGCHI Conference on Human Factors in Computing Systems - CHI '13, Paris. pp. 3307–3316. Association for Computing Machinery, New York (2013). https://doi.org/10/gj44p8.
12. Böckle, M., Novak, J., Bick, M.: Exploring Gamified Persuasive System Design for Energy Saving. Journal of Enterprise Information Management. 33, 1337–1356 (2020). https://doi.org/10/gj44qv.
13. Liu, D., Santhanam, R., Webster, J.: Toward Meaningful Engagement: A Framework for Design and Research of Gamified Information Systems. MIS Quarterly. 41, 1011–1034, A1–A4 (2017). https://doi.org/10/gdsh7x.
14. Robson, K., Plangger, K., Kietzmann, J., McCarthy, I., Pitt, L.: Is It All a Game? Understanding the Principles of Gamification. Business Horizons. 58, 411–420 (2015). https://doi.org/10/gfgp2j.
15. Whitson, J.: Gaming the Quantified Self. Surveillance & Society. Vol 11, 163–176 (2013). https://doi.org/10/gk9wf4.
16. Oinas-Kukkonen, H., Harjumaa, M.: A Systematic Framework for Designing and Evaluating Persuasive Systems. In: Oinas-Kukkonen, H., Hasle, P., Harjumaa, M., Segerståhl, K., and Øhrstrøm, P. (eds.) Persuasive Technology. PERSUASIVE 2008, Oulu. pp. 164–176. Springer, Cham (2008). https://doi.org/10/b9kzsm.
17. Chiu, M.-C., Kuo, T.-C., Liao, H.-T.: Design for Sustainable Behavior Strategies: Impact of Persuasive Technology on Energy Usage. Journal of Cleaner Production. 248, art. 119214 (2020). https://doi.org/10/gghpv6.
18. Hamari, J., Koivisto, J.: Why Do People Use Gamification Services? International Journal of Information Management. 35, 419–431 (2015). https://doi.org/10/f7nnxm.
19. Johnston, A., Warkentin, M.: Fear Appeals and Information Security Behaviors: An Empirical Study. MIS Quarterly. 34, 549–566 (2010). https://doi.org/10/ggf5p5.
20. Johnson, D., Horton, E., Mulcahy, R., Foth, M.: Gamification and Serious Games within the Domain of Domestic Energy Consumption: A Systematic Review. Renewable and Sustainable Energy Reviews. 73, 249–264 (2017). https://doi.org/10/f99frw.
21. Lehrer, D., Vasudev, J.: Evaluating a Social Media Application for Sustainability in the Workplace. In: Karat, C.-M., Lund, A., Coutaz, J., and Karat, J. (eds.) Proceedings of the SIGCHI Conference on Human Factors in Computing Systems - CHI '11, Vancouver. pp. 2161–2166. Association for Computing Machinery, New York (2011). https://doi.org/10/djgg4s.
22. Lehrer, D., Vasudev, J., Kaam, S.: A Usability Study of a Social Media Prototype for Building Energy Feedback and Operations. Proceedings 2014 ACEEE Summer Study on Energy Efficiency in Buildings. 173–186 (2014).
23. Deriu, J., Rodrigo, A., Otegi, A., Echegoyen, G., Rosset, S., Agirre, E., Cieliebak, M.: Survey on Evaluation Methods for Dialogue Systems. Artificial Intelligence Review. 54, 755–810 (2021). https://doi.org/10/gg5t8r.
24. Are "Assume" and "Presume" Synonyms?, https://www.merriam-webster.com/words-at-play/assume-vs-presume, last accessed 2021/06/09.

25. Kumar, J.: Gamification at Work: Designing Engaging Business Software. In: Marcus, A. (ed.) Design, User Experience, and Usability. Health, Learning, Playing, Cultural, and Cross-Cultural User Experience. DUXU 2013, Part II, Las Vegas. pp. 528–537. Springer, Berlin (2013). https://doi.org/10/ghm2qq.
26. Makanawala, P., Godara, J., Goldwasser, E., Le, H.: Applying Gamification in Customer Service Application to Improve Agents' Efficiency and Satisfaction. In: Marcus, A. (ed.) Design, User Experience, and Usability. Health, Learning, Playing, Cultural, and Cross-Cultural User Experience. DUXU 2013, Part II, Las Vegas. pp. 548–557. Springer, Berlin (2013). https://doi.org/10/gk96cm.
27. Fritz, T., Huang, E., Murphy, G., Zimmermann, T.: Persuasive Technology in the Real World: A Study of Long-Term Use of Activity Sensing Devices for Fitness. In: Proceedings of the SIGCHI Conference on Human Factors in Computing Systems - CHI '14, Toronto. pp. 487–496. Association for Computing Machinery, New York (2014). https://doi.org/10/bbdv.
28. Lockton, D., Nicholson, L., Cain, R., Harrison, D.: Persuasive Technology for Sustainable Workplaces. Interactions. 21, 58–61 (2014). https://doi.org/10/gk9wd8.
29. Chesney, T., Chuah, S.-H., Hoffmann, R., Larner, J.: The Influence of Influence: The Effect of Task Repetition on Persuaders and Persuadees. Decision Support Systems. 94, 12–18 (2017). https://doi.org/10/gj44qw.
30. Stieglitz, S., Potthoff, T., Kißmer, T.: Digital Nudging am Arbeitsplatz: Ein Ansatz zur Steigerung der Technologieakzeptanz. HMD Praxis der Wirtschaftsinformatik. 54, 965–976 (2017). https://doi.org/10/gc8stw.
31. Mogles, N., Padget, J., Gabe-Thomas, E., Walker, I., Lee, J.: A Computational Model for Designing Energy Behaviour Change Interventions. User Modeling and User-Adapted Interaction. 28, 1–34 (2018). https://doi.org/10/gc66bp.
32. Khashe, S., Lucas, G., Becerik-Gerber, B., Gratch, J.: Establishing Social Dialog between Buildings and Their Users. International Journal of Human-Computer Interaction. 35, 1545–1556 (2019). https://doi.org/10/gj44qc.
33. Khataei, S., Hine, M., Arya, A.: The Design, Development and Validation of a Persuasive Content Generator. Journal of International Technology and Information Management. 29, 46–80 (2021).
34. Steelman, Z., Hammer, B., Limayem, M.: Data Collection in the Digital Age: Innovative Alternatives to Student Samples. MIS Quarterly. 38, 355–378, A1–A20 (2014). https://doi.org/10/gghqfj.
35. Gupta, A., Kannan, K., Sanyal, P.: Economic Experiments in Information Systems. MIS Quarterly. 42, 595–606 (2018). https://doi.org/10/gdsh6k.
36. Schmid, Y., Dowling, M.: New Work: New Motivation? A Comprehensive Literature Review on the Impact of Workplace Technologies. Management Review Quarterly. (2020). https://doi.org/10/gkmwv4.